\documentclass[conference]{IEEEtran}
\IEEEoverridecommandlockouts

\usepackage[utf8]{inputenc}
\usepackage[final]{microtype}
\usepackage{amsmath,amssymb,amsfonts}
\usepackage{amsthm}
\usepackage{algorithm}
\usepackage{algorithmicx}
\usepackage{siunitx}
\usepackage{graphicx}
\usepackage{textcomp}
\usepackage[hidelinks]{hyperref}
\usepackage{xcolor}
\usepackage{bm}
\usepackage{mathrsfs}
\usepackage{physics}
\usepackage{multirow}
\usepackage[compress]{cite}

\usepackage{tikz}
\usepackage{pgfplots}
\usepackage{pgfgantt}
\usepackage{pdflscape}
\usepackage{grffile}
\pgfplotsset{compat=newest}
\pgfplotsset{plot coordinates/math parser=false}

\def\BibTeX{{\rm B\kern-.05em{\sc i\kern-.025em b}\kern-.08em
    T\kern-.1667em\lower.7ex\hbox{E}\kern-.125emX}}

\theoremstyle{plain}
\newtheorem{thm}{Theorem}
\newtheorem{lem}[thm]{Lemma}

\theoremstyle{definition}
\newtheorem{defn}{Definition}

\theoremstyle{remark}
\newtheorem{rem}{Remark}

\begin{document}

\title{Towards Data-Driven Multi-Stage OPF
	\thanks{OM and TF are with the Institute of Control Systems, Hamburg University of Technology, Hamburg, Germany. OM and TF acknowledge funding in the course of TRR 391 \textit{Spatio-temporal Statistics for the Transition of Energy and Transport} (520388526) by the Deutsche Forschungsgemeinschaft (DFG, German Research Foundation). E-mail OM and TF: oleksii.molodchyk@tuhh.de, timm.faulwasser@ieee.org.}
	\thanks{AE is with logarithmo GmbH \& Co. KG, Dortmund, Germany. PS and KW are with the Optimization-based Control Group, TU Ilmenau, Germany. PS is grateful for the support from the Carl Zeiss Foundation (VerneDCt – Project No. 2011640173).}
}

\author{Oleksii Molodchyk, Philipp Schmitz, Alexander Engelmann, Karl Worthmann, and Timm Faulwasser}

\maketitle

\begin{abstract}
	The operation of large-scale power systems is usually scheduled ahead via numerical optimization.
	However, this requires models of grid topology, line parameters, and bus specifications.
	Classic approaches first identify the network topology, i.e., the graph of interconnections and the associated impedances.
	The power generation schedules are then computed by solving a multi-stage optimal power flow (OPF) problem built around the model.
	In this paper, we explore the prospect of data-driven approaches to multi-stage optimal power flow.
	Specifically, we leverage recent findings from systems and control to bypass the identification step and to construct the optimization problem directly from data.
	We illustrate the performance of our method on a 118-bus system and compare it with the classical identification-based approach.
\end{abstract}


\section{Introduction}
Electrical power systems are becoming increasingly more complex due to increasing share of renewable energy sources and couplings with other sectors such as, e.g., heating, gas, and transportation.
Hence, optimization-based methods for power system operation and control are of growing research interest~\cite{Garces2021,Capitanescu2011,Kourounis2018,Levron2013,Faulwasser2019}.
These approaches built upon grid models. Classic approaches  identify models in a sequential procedure usually consisting of two steps: i) topology detection, ii) parameter estimation, cf. \cite[Chap 6.1]{Monticelli2012}, \cite{Li2013,Donmez2022,Farajollahi2020,deka2022learning,Du2020}.
Although working usually well in practice, these approaches are quite complex and involve several steps, each of which has to be implemented and tuned for numerical performance.

As an alternative, data-driven approaches are increasingly popular for control of dynamic systems as they mitigate system identification and as they directly work on measured data.
A large swath of recent research on data-driven control relies a result by Willems and co-authors called the fundamental lemma, cf.~\cite{Willems2005} for the original reference and \cite{Coulson2019,Dorfler2023,Markovsky2021,Verheijen2023} for recent influential works.
These approaches have, e.g., been considered for frequency control and power oscillation damping in power systems \cite{Markovsky2022,Huang2022,Schmitz2022b}.
For the operation of power systems including storage and considering time scales in the order of several minutes, i.e., for multi-stage power flow problems in the context of quasi-stationary operation, data-driven methods have so far not been considered.

In the present paper, we develop a purely data-driven predictive control scheme for multi-stage optimal power flow (OPF) with battery energy storage systems (BESS), which avoids any explicit model identification and parametrization.
To this end, we take a descriptor systems perspective which allows for the explicit consideration of algebraic constraints (the network model)~\cite{Schmitz2022a}.
We explore the close connections of our data-based approach to existing regression-based grid identification techniques \cite{Chen2014, Ardakanian2019,Liu2019,Yuan2023}.
We illustrate the numerical performance of our method on a 118-bus system example.

\section{Preliminaries and Problem Statement}
Consider a balanced electric power system at steady state with buses indexed by a set $\mathcal{N}$. It is equipped with controllable generators and BESS labeled by $\mathcal{G}$; the subset $\mathcal{S} \subset \mathcal{G}$ labels only BESS. Similarly, uncontrollable demands are indexed by the set $\mathcal{D}$. Without loss of generality, we define $1 \in \mathcal{N}$ as the reference bus; it accommodates the slack generator $1 \in \mathcal{G}$ which compensates for the difference between total generation and load in the network.
Transmission lines, cables, and transformers are represented via the edge set $\mathcal{E} \subset \mathcal{N} \times \mathcal{N}$, wherein each edge $(i,j) \in \mathcal{E}$ is endowed with reactance $x_{ij} = x_{ji}>0$ and transformer tap ratio $\tau_{ij} = \tau_{ji}>0$.
The corresponding bus susceptance matrix $\mathbf{B} \in \mathbb{R}^{\abs{\mathcal{N}} \times \abs{\mathcal{N}}}$ reads
\begin{equation} \label{eq:Bbus}
	(\mathbf{B})_{ij} \doteq
	\begin{cases}
		\sum\limits_{k \neq i } \frac{1}{x_{ik} \tau_{ik}}, & i = j,    \\
		-\frac{1}{x_{ij} \tau_{ij}},                        & i \neq j.
	\end{cases}
\end{equation}

For the sake of simplicity, we consider the DC setting, which relies on a set of assumptions such as unit voltages, small phase angles, and lossless lines~\cite{Stott2009, Frank2016}. For the  DC model, the vector of net active power injections (i.e., the differences of generation and consumption at all nodes) and the voltage phasors $\theta \in \mathbb{R}^{\abs{\mathcal{N}}}$, are related via
\begin{equation}
	\label{eqn:dc_pf}
	\mathbf{C}_\mathrm{g} p(k) - \mathbf{C}_\mathrm{d} w(k) = \mathbf{B} \theta(k),
\end{equation}
at every time step $k \in \mathbb{N}_0$. Here, $p(k) =[g(k)^\top\; s(k)^\top]^\top \in \mathbb{R}^{\abs{\mathcal{G}}}$
is the input vector consisting of active generator power $g(k) \in \mathbb{R}^{\abs{\mathcal{G}}-\abs{\mathcal{S}}}$ and storage power $s(k) \in \mathbb{R}^{\abs{\mathcal{S}}}$.
Uncontrollable active power demands are collected in $w(k) \in \mathbb{R}^{\abs{\mathcal{D}}}$.
The matrix $\mathbf{C}_\mathrm{g} \in \lbrace 0,1 \rbrace^{\abs{\mathcal{N}} \times \abs{\mathcal{G}}}$ maps the generator/storage and, respectively, the demand power to the buses in $\mathcal{N}$ via
\begin{equation*}
	(\mathbf{C}_\mathrm{g})_{ij} \doteq
	\begin{cases}
		\displaystyle
		1, & \text{if } j \in \mathcal{G} \text{ is connected to bus } i \in \mathcal{N}, \\
		0, & \text{otherwise}.
	\end{cases}
\end{equation*}
The matrix $\mathbf{C}_\mathrm{d} \in \lbrace 0,1 \rbrace^{\abs{\mathcal{N}} \times \abs{\mathcal{D}}}$ is defined analogously to the above; in the same fashion, we define $\mathbf{C}_{\mathrm s}$ selecting the BESS via $s(k) = \mathbf{C}_\mathrm{s} p(k)$. Additionally, we construct the branch susceptance matrix ${\mathbf B}_\mathrm{f}$, which maps the voltage angles $\theta(k)$ to a vector $f(k) \in \mathbb{R}^{|\mathcal E|}$ of active power flows over the branches in $\mathcal{E}$ via $f(k) = {\mathbf B}_\mathrm{f} \,\theta(k)$.

We consider multi-stage power systems operation assuming ideal (lossless) BESS.
Thus a simple model reads
\begin{align} \label{eq:storDyn}
	e(k+1) = e(k) - \delta s(k), \qquad e(0) \doteq e^0,
\end{align}
where $e \in \mathbb R^{\abs{\mathcal{S}}}$ is the vectorized state of charge, $s \in \mathbb R^{\abs{\mathcal{E}}}$ are BESS feed-ins, $e^0\in \mathbb R^{\abs{\mathcal{E}}}$ is the vectorized initial state of charge, and $\delta>0$ is the sampling period.
We collect the slack generation $p_1(k)$, power flows $f(k)$, and stored energy $e(k)$ in the (output) vector $y(k) \doteq [p_1(k)^\top\;\;f(k)^\top\;\;e(k)^\top]^\top$. We define the (input) vector as $u(k) \doteq p_{\mathcal{N} \setminus 1}(k)$, collecting all generations, except the one of the slack generator.

Now we are ready to state a deterministic, multi-stage DC OPF problem over the prediction horizon $L \in \mathbb{N}$
\begin{subequations}
	\label{eqn:dc_pf_mpc}
	\begin{equation}
		\operatorname*{minimize}_{\{(u(k),y(k),\theta(k))\}_t^{t+L-1}} \ \sum_{k=t}^{t+L-1} \ell(u(k),y(k))
	\end{equation}
	subject to for all $k \in \{ t, \cdots, t+L-1 \}$
	\begin{align}
		\label{eqn:lti_bess}
		e(k+1)                                                      & = e(k) - \delta \mathbf{C}_{\mathrm s} p(k),\ e(t) = e^0 \\
		\mathbf{C}_{\mathrm{g}} p(k) - \mathbf{C}_{\mathrm{d}} w(k) & = \mathbf B \theta(k),  \; \theta_1(k) = 0               \\
		\label{eqn:p_flows}
		f(k)                                                        & = {\mathbf B}_\mathrm{f} \,\theta(k),                    \\
		\label{eq:setCstr}
		u(k)                                                        & \in \mathbb{U}, \; y(k) \in \mathbb{Y},
	\end{align}
\end{subequations}
with stage cost function $\ell : \mathbb{R}^{n_u} \times \mathbb{R}^{n_y} \to \mathbb{R}$ and constraint sets $\mathbb{U} \subseteq \mathbb{R}^{\abs{\mathcal{G}}-1}$ and $\mathbb{Y} \subseteq \mathbb{R}^{ \abs{\mathcal{S}} + \abs{\mathcal{E}} + 1}$.

Note that the susceptance matrix ${\mathbf B}$ given by \eqref{eq:Bbus} is the graph Laplacian corresponding to a weighted network graph. Therefore, zero is an eigenvalue of ${\mathbf B}$ with multiplicity equal to the number of connected components in the network.\footnote{In our case, the multiplicity of a zero eigenvalue is equal to one because we aim to control only one connected electric network.} To guarantee the linear independence of rows in \eqref{eqn:dc_pf}, we remove the row corresponding to the slack bus, and ---because the reference node is fixed at $\theta_1 = 0$ ---the first column associated with the reference bus~1.
This yields the reduced matrix $\tilde{\mathbf B}$.
The matrices $\tilde{\mathbf B}_\mathrm{f}$, $\tilde{\mathbf{C}}_{\mathrm s}$, $\tilde{\mathbf{C}}_{\mathrm g}$, and $\tilde{\mathbf{C}}_{\mathrm d}$ are obtained in a similar fashion. We also introduce $\tilde \theta \doteq \theta_{\mathcal{N}\setminus 1}$.

Equations \eqref{eqn:lti_bess}-\eqref{eqn:p_flows} can be regarded as a linear time-invariant descriptor system, which consists of dynamic and algebraic parts.
These equations are summarized in matrix-vector form in \eqref{eqn:dc_pf_lds} where the DC power flow equations \eqref{eqn:dc_pf} are solved explicitly by inverting $\tilde{\mathbf B}$. In the following, we discuss how linear descriptor systems can be controlled in a data-driven fashion relying on the behavioral theory \cite{Polderman1998}.

\begin{figure*}[!t]
	\normalsize
	\begin{subequations}
		\label{eqn:dc_pf_lds}
		\begin{alignat}{1}
			\label{eqn:dc_pf_lds1}
			\begin{bmatrix}
				\mathbf{I}_{\abs{\mathcal{S}}} & \mathbf{0} \\
				\mathbf{0}                     & \mathbf{0}
			\end{bmatrix}
			\begin{bmatrix}
				e(k+1) \\
				\tilde{ \theta}(k+1)
			\end{bmatrix}   & =
			\begin{bmatrix}
				\mathbf{I}_{\abs{\mathcal{S}}} & 0                                \\
				0                              & \mathbf{I}_{\abs{\mathcal{N}}-1}
			\end{bmatrix}
			\begin{bmatrix}
				e(k) \\
				{\tilde  \theta}(k)
			\end{bmatrix} +
			\begin{bmatrix}
				-\delta \tilde{\mathbf{C}}_{\mathrm s} \\
				-\tilde{\mathbf B}^{-1} \tilde{\mathbf{C}}_{\mathrm g}
			\end{bmatrix} u(k) +
			\begin{bmatrix}
				\mathbf{0}                                            \\
				\tilde{\mathbf B}^{-1} \tilde{\mathbf{C}}_{\mathrm d} \\
			\end{bmatrix} w(k), \\
			\label{eqn:dc_pf_lds2}
			y(k) = \begin{bmatrix}
				       e(k)   \\
				       p_1(k) \\
				       f(k)
			       \end{bmatrix} & =
			\begin{bmatrix}
				\mathbf{I}_{\abs{\mathcal{S}}} & \mathbf{0}                   \\
				\mathbf{0}                     & \mathbf{0}                   \\
				\mathbf{0}                     & \tilde{\mathbf B}_\mathrm{f}
			\end{bmatrix}
			\begin{bmatrix}
				e(k) \\
				\tilde{ \theta}(k)
			\end{bmatrix} +
			\begin{bmatrix}
				\mathbf{0}       \\
				-\mathbf{1}^\top \\
				\mathbf{0}
			\end{bmatrix}u(k) +
			\begin{bmatrix}
				\mathbf{0}      \\
				\mathbf{1}^\top \\
				\mathbf{0}
			\end{bmatrix} w(k).
		\end{alignat}
	\end{subequations}

	\hrulefill
	\vspace*{4pt}
\end{figure*}

\subsection{Linear descriptor systems} \label{sec:LDS}
Observe that the descriptor system \eqref{eqn:dc_pf_lds} is written in a quasi-Weierstraß form
\begin{subequations} \label{eqn:lds_tr}
	\begin{align}
		x_1(t+1)  & =  A_1 x_1(t) + B_1  u(t) + F_1 w(t),  \label{eqn:lds_tr_a} \\
		Nx_2(t+1) & = x_2(t) + B_2 u(t) + F_2 w(t), \notag                      \\
		y(t)      & = \begin{bmatrix} C_1 & C_2 \end{bmatrix}
		\begin{bmatrix} x_1(t) \\ x_2(t) \end{bmatrix} + D u(t) + Gw(t).
	\end{align}
\end{subequations}
Here, $\bm{q}, \bm{r} \in \mathbb{N}$, $A_1 \in \mathbb{R}^{\bm{q} \times \bm{q}}$, and $N \in \mathbb{R}^{\bm{r} \times \bm{r}}$ is nilpotent with nilpotency index $\bm{s} \in \mathbb{N}$, i.e. $N^{\bm s}=0$ and $N^{\bm s - 1} \neq 0$.

The dynamic part \eqref{eqn:lds_tr_a} corresponding to $x_1$ models the storages and is contained in the  subsystem $(A_1, B_1, F_1, C_1)$. The algebraic power balance equations \eqref{eqn:dc_pf} are expressed via $(N, B_2, F_2, C_2)$ and can be solved for the state $x_2$. Since  \eqref{eqn:dc_pf_lds} is already given in quasi-Weierstrass form \eqref{eqn:lds_tr}, it can readily be examined in terms of the systems and control notions R-controllability and R-observability as defined below.

\begin{defn}[R-controllability and R-observability {\cite{dai.1989}}]
	\label{def:r_ctrl_r_obsr}
	System \eqref{eqn:lds_tr} is called \emph{R-controllable} if $\rank([(A_1-\lambda \mathbf{I}_{\bm{q}}) \;B_1 \; F_1])=\bm{q}$ for all $\lambda\in\mathbb C$, and \emph{R-observable} if $\rank([(A_1^\top-\lambda \mathbf{I}_{\bm{q}}) \;C_1^\top ])=\bm{q}$ for all $\lambda\in\mathbb C$.
\end{defn}
It is worth noting that, by definition, $\rank \tilde{\mathbf{C}}_{\mathrm s} = \rank C_1 = \abs{\mathcal{S}} = \bm{q}$. This allows to conclude that \eqref{eqn:dc_pf_lds} is both R-controllable and R-observable. As we see below, this renders data-driven predictive control methods applicable.

\subsection{Data-driven Predictive Control}
Given an abstract signal sequence $\mathbf v = (v(0), v(1),\dots)$ in $\mathbb R^{n_\mathrm v}$ we denote the restriction of $\mathbf{v}$ to the time window $[t:T] \doteq \{t,\dots, T\}$ by $\mathbf{v}_{[t:T]}$ and for $N\in \mathbb N$ we consider the Hankel matrix
\begin{equation*}
	\mathscr{H}_N\left(\mathbf{v}_{[t:T]}\right) \doteq
	\begin{bmatrix}
		v(t)     & \cdots & v(T-N+1) \\
		\vdots   & \ddots & \vdots   \\
		v(t+N-1) & \cdots & v(T)
	\end{bmatrix}.
\end{equation*}
The sequence $\mathbf{v}_{[t:T]}$ is called \emph{persistently exciting of order $N$}, if $\rank \left( \mathscr{H}_N\left(\mathbf{v}_{[t:T]}\right) \right) = n_{\mathrm v} N$.

Next we present a data-driven parametrization of all input-output trajectories of \eqref{eqn:lds_tr} which is inspired by the original results of Willems et al.~\cite{Willems2005}.\footnote{The analysis of linear time-invariant systems does not require the need to consider descriptor system, cf.~\cite{willems86i,Willems91}---provided one is free to assign inputs and outputs without limits set by applications. From a power systems perspective, however, inputs and outputs are clearly defined by their physical meaning. Hence there is the need to consider the descriptor setting in detail.}
\begin{lem}[{\cite[Lemma~2]{Schmitz2022a}}]
	\label{lem:lds_fundamental}
	Assume that \eqref{eqn:lds_tr} is R-controllable. Let $(\mathbf{u}^{\mathrm d}, \mathbf{w}^{\mathrm d}, \mathbf{y}^{\mathrm d})_{[1:T]}$ be an offline-recorded length-$T$ input-output trajectory of system~\eqref{eqn:lds_tr} such that the joint input signal $\left(\mathbf{u}^{\mathrm d}, \mathbf{w}^{\mathrm d}\right)_{[1:T]}$ is persistently exciting of order $N + \bm{q} + \bm{s} - 1$, where $N\in\mathbb N$. Then, $(\mathbf{u}, \mathbf{w}, \mathbf{y})_{[1:N]}$ is an input-output trajectory of system~\eqref{eqn:lds_tr} if and only if there exists $\alpha \in \mathbb{R}^{T-N-\bm{s} +2}$ such that
	\begin{equation}
		\label{eq:hankel_rep}
		\begin{bmatrix}
			\mathscr{H}_N(\mathbf{u}^{\mathrm d}_{[1:T-\bm{s}+1]}) \\
			\mathscr{H}_N(\mathbf{w}^{\mathrm d}_{[1:T-\bm{s}+1]}) \\
			\mathscr{H}_N(\mathbf{y}^{\mathrm d}_{[1:T-\bm{s}+1]})
		\end{bmatrix} \alpha =
		\begin{bmatrix}
			\mathbf{u}_{[1:N]} \\
			\mathbf{w}_{[1:N]} \\
			\mathbf{y}_{[1:N]}
		\end{bmatrix}.
	\end{equation}
\end{lem}
Lemma~\ref{lem:lds_fundamental} allows for non-parametric representations of discrete-time LTI systems based purely on the input-output trajectories. This can be exploited to build a data-driven predictive controller. In order to maintain consistency with the systems' dynamics, alignment of the latent state variable at the junction of a past trajectory and its future prediction is crucial.
Although the representation~\eqref{eq:hankel_rep} of input-output trajectories does not allow direct access to the internal state variable, R-observability of the system ensures that the output is uniquely determined given an input-output trajectory over a sufficiently long horizon $N\geq T_\text{ini}\doteq \bm{q}+\bm{s}-1$, cf.~\cite[Lemma~1]{Schmitz2022a}.

Consider a prediction horizon $L\in\mathbb N$ and choose $N=L+T_\text{ini}$. Let $(\mathbf{u}^{\mathrm d}, \mathbf{w}^{\mathrm d}, \mathbf{y}^{\mathrm d})_{[1:T]}$ be offline-recorded input-output data such that $(\mathbf{u}^{\mathrm d}, \mathbf{w}^{\mathrm d})_{[1:T]}$ is persistently exciting of order $N+\bm{q}+\bm{s}-1 = L+2(\bm{q}+\bm{s})-2$. Note that the persistency of excitation order enforces a lower bound on the data demand, $T\geq (n_\mathrm{u}+n_\mathrm{w}+1)(L+2(\bm q +\bm s -1))-1$. We separate the Hankel matrices from (\ref{eq:hankel_rep}) into past and future parts denoted by the respective subscripts ``P'' and ``F'' into
\begin{align}
	\label{eqn:past_and_future_hankel}
	 & \mathscr{H}_{T_{\mathrm{ini}} + L}(\mathbf{z}^{\mathrm d}_{[1:T-s+1]}) = \begin{bmatrix}
		                                                                            \mathscr{H}_{T_{\mathrm{ini}}}(\mathbf{z}^{\mathrm d}_{[1:T-L-s+1]}) \\
		                                                                            \mathscr{H}_{L}(\mathbf{z}^{\mathrm d}_{[T_\mathrm{ini}+1:T-s+1]})
	                                                                            \end{bmatrix}=
	\begin{bmatrix}
		Z_{\mathrm P} \\
		Z_{\mathrm F}
	\end{bmatrix},
\end{align}
which is implemented with respect to the input $u$, the output $y$, and the exogeneous disturbance $w$, i.e., for all $(z,Z) \in \{(u,U), (y,Y), (w,W)\}$.

We apply \eqref{eqn:past_and_future_hankel} to formulate the data-driven optimization problem. Given the measured past (initial) input-output trajectory $(\mathbf{u}, \mathbf{w}, \mathbf{y})_{[t-T_{\mathrm{ini}}:t-1]}$, we consider the discrete-time optimal control problem (OCP)
\begin{subequations}
	\label{eqn:dd_ocp}
	\begin{equation}
		\operatorname*{minimize}_{\hat{\mathbf{u}}, \hat{\mathbf{y}},\alpha} \ \sum_{k=t}^{t+L-1} \ell(\hat{u}(k),\hat{y}(k)) \label{eqn:dd_costs}
	\end{equation}
	subject to
	\begin{equation}
		\begin{bmatrix}
			U_{\mathrm F} \\
			W_{\mathrm F} \\
			Y_{\mathrm F}
		\end{bmatrix} \alpha =
		\begin{bmatrix}
			\hat{\mathbf{u}}_{[t:t+L-1]} \\
			\hat{\mathbf{w}}_{[t:t+L-1]} \\
			\hat{\mathbf{y}}_{[t:t+L-1]}
		\end{bmatrix}, \quad  \begin{bmatrix}
			U_{\mathrm P} \\
			W_{\mathrm P} \\
			Y_{\mathrm P}
		\end{bmatrix} \alpha =
		\begin{bmatrix}
			\mathbf{u}_{[t-T_{\mathrm{ini}}:t-1]} \\
			\mathbf{w}_{[t-T_{\mathrm{ini}}:t-1]} \\
			\mathbf{y}_{[t-T_{\mathrm{ini}}:t-1]}
		\end{bmatrix}\label{eqn:dd_sys}
	\end{equation}
	and for all $k \in \{ t, \cdots, t+L-1 \}$
	\begin{align}
		 & u^\mathrm{min} \leq \hat{u}(k) \leq u^\mathrm{max}, \label{eqn:dd_box_u} \\
		 & \hat{w}(k) = w(k), \label{eqn:dd_demand}                                 \\
		 & y^\mathrm{min} \leq \hat{y}(k) \leq y^\mathrm{max} \label{eqn:dd_box_y}.
	\end{align}
\end{subequations}
Based on \eqref{eqn:dd_ocp}, we can formulate a data-driven predictive control scheme. Its overall workflow is summarized in Algorithm~\ref{alg:deepc}. Steps~1-4 describe the offline collection of data needed to construct a non-parametric system representation \eqref{eq:hankel_rep}.
Steps~5-9 refer to the actual closed-loop control.
Algorithm~\ref{alg:deepc} allows for system \eqref{eqn:dc_pf_lds} to be controlled using the same constraints and stage costs as in \eqref{eqn:dc_pf_mpc} but without the need for a state-space representation \eqref{eqn:lds_tr}. The only model-related information that is required is the knowledge of $\bm{q}$, $\bm{s}$ (or at least upper bounds).
\begin{algorithm}
	\textbf{Input:} R-controllable and R-observable descriptor system \eqref{eqn:lds_tr}; stage cost $\ell : \mathbb{R}^{n_{\mathrm u}} \times \mathbb{R}^{n_{\mathrm y}} \rightarrow \mathbb{R}$; constraints $\mathbb{U}$, $\mathbb{Y}$; prediction and control horizon $L, L_c \in \mathbb N$; initialization trajectory length $T_{\mathrm{ini}} \doteq \bm{q} + \bm{s} - 1$.
	\begin{algorithmic}[1]
		\State Predict disturbances $\mathbf{w}^{\mathrm d}_{[1:T]}$ and generate inputs $\mathbf{u}^{\mathrm d}_{[1:T]}$ such that $(\mathbf{u}^{\mathrm d},\mathbf{w}^{\mathrm d})_{[1:T]}$ is persistently exciting of order $T_{\mathrm{ini}} + L + \bm{q} + \bm{s} - 1$. \label{itm:deepc_1}
		\State Apply $\mathbf{u}^{\mathrm d}_{[1:T]}$ to (\ref{eqn:lds_tr}) and get $(\mathbf{u}^{\mathrm d},\mathbf{w}^{\mathrm d},\mathbf{y}^{\mathrm d})_{[1:T]}$. \label{itm:deepc_2}
		\State Assemble $U_{\mathrm P}$, $W_{\mathrm P}$, $Y_{\mathrm P}$, $U_{\mathrm F}$, $W_{\mathrm F}$, and $Y_{\mathrm F}$ from $(\mathbf{u}^{\mathrm d},\mathbf{w}^{\mathrm d},\mathbf{y}^{\mathrm d})_{[1:T]}$ using (\ref{eqn:past_and_future_hankel}). \label{itm:deepc_3}
		\State Set $t \doteq T + 1$. \label{itm:deepc_4}
		\State Retrieve measured trajectory $(\mathbf{u}, \mathbf{w}, \mathbf{y})_{[t-T_{\mathrm{ini}}:t-1]}$. \label{itm:deepc_5}
		\State Predict future disturbance $\mathbf{w}_{[t:t+L-1]}$
		\State Solve (\ref{eqn:dd_ocp}) for $\hat{\mathbf{u}}^\star = U_{\mathrm F} \alpha^\star$. \label{itm:deepc_6}
		\State Apply control sequence $\hat{\mathbf{u}}^\star_{[t:t+L_c-1]}$ to (\ref{eqn:lds_tr}). \label{itm:deepc_7}
		\State Set $t \doteq t + L_c$. \label{itm:deepc_8}
		\State Go to Step~5. \label{itm:deepc_9}
	\end{algorithmic}
	\caption{Data-driven predictive control}
	\label{alg:deepc}
\end{algorithm}

\section{Data-Driven Multi-Stage OPF}
We explore the applicability of the data-driven predictive control scheme from the previous section to the power system model  \eqref{eqn:dc_pf_lds}.
The system matrices of the quasi-Weierstrass form \eqref{eqn:lds_tr} are
\begin{subequations}
	\label{eqn:lds_matrices}
	\begin{alignat}{2}
		 & A_1 = \phantom{-}\mathbf{I}_{\abs{\mathcal{S}}},               &  & \qquad N = \mathbf{0}_{ (\abs{\mathcal{N}}-1) \times (\abs{\mathcal{N}}-1)},
		\label{eqn:matrix_def}                                                                                                                              \\
		 & B_1 = -\delta \tilde{\mathbf{C}}_{\mathrm s},                  &  & \qquad F_1 = \mathbf{0}_{\abs{\mathcal{S}} \times \abs{\mathcal{D}}},        \\
		 & B_2 = -\tilde{\mathbf{B}}^{-1} \tilde{\mathbf{C}}_{\mathrm g}, &  & \qquad
		F_2 = \tilde{\mathbf{B}}^{-1} \tilde{\mathbf{C}}_{\mathrm d},                                                                                       \\
		 & C_1 = \begin{bmatrix}
			         \mathbf{I}_{\abs{\mathcal{S}}} \\
			         \mathbf{0}                     \\
			         \mathbf{0}
		         \end{bmatrix},                  &  & \qquad
		C_2 = \begin{bmatrix}
			      \mathbf{0} \\
			      \mathbf{0} \\
			      \tilde{\mathbf{B}}_\mathrm{f}
		      \end{bmatrix}.\label{eqn:Cdef}
	\end{alignat}
\end{subequations}
System-theoretically speaking, a battery is a scalar and causal dynamic system, hence, $\bm{s} = 1$.
The dimension of the storage dynamics corresponds to the number of the storages, thus $\bm{q} = \abs{\mathcal{S}}$. The joint dimension of inputs and disturbances is given by $n_\mathrm u + n_\mathrm w = \abs{\mathcal G}-1 + \abs{\mathcal D}$. These specifications can be derived from the definitions of \autoref{sec:LDS}, \eqref{eqn:matrix_def}, and \eqref{eqn:Cdef}.
Thus, the initial trajectory length should satisfy $T_{\mathrm{ini}} = 1$.
Meanwhile, the offline (training) trajectory length is lower-bounded by $T \geq (\abs{\mathcal{G}} + \abs{\mathcal D})(L + 2\abs{\mathcal{S}}) - 1$, where $L$ is the prediction horizon.

\begin{rem}[Simplifying the non-parametric system representation~\eqref{eqn:dd_sys}]
	\label{rem:lin_dd_sim}
	Due to the static nature of the DC power flow equations, the trajectory of past disturbances $\mathbf{w}_{[t-T_\mathrm{ini} : t-1]}$ has no influence on the future outputs of the system \eqref{eqn:dc_pf_lds}. In fact, the dynamic part of \eqref{eqn:dc_pf_lds} stems only from the BESS equations \eqref{eqn:lti_bess}. Therefore, \eqref{eqn:dd_sys} can be simplified to
	\begin{equation}
		\label{eqn:dd_simplified}
		\begin{bmatrix}
			S_{\mathrm P} \\
			E_{\mathrm P} \\
			U_{\mathrm F} \\
			W_{\mathrm F} \\
			Y_{\mathrm F}
		\end{bmatrix} \alpha =
		\begin{bmatrix}
			\mathbf{s}_{[t-T_\mathrm{ini} : t-1]} \\
			\mathbf{e}_{[t-T_\mathrm{ini} : t-1]} \\
			\hat{\mathbf{u}}_{[t:t+L-1]}          \\
			\hat{\mathbf{w}}_{[t:t+L-1]}          \\
			\hat{\mathbf{y}}_{[t:t+L-1]}
		\end{bmatrix}.
	\end{equation}
	Here, following the previous remark, the initial trajectory only entails the past measurements of the powers and states of charge of the BESS. Therefore, the Hankel matrices with subscript ``P" corresponding to the initial trajectory are $S_{\mathrm P}$ and $E_{\mathrm P}$ containing the BESS power and state of charge data, respectively.

	The complexity of the problem \eqref{eqn:dd_ocp} can be additionally reduced upon introducing the substitution
	\begin{equation}
		\label{eqn:substitution}
		\alpha = W_{\mathrm F}^\dagger \mathbf{w}_{[t:t+L-1]} + \operatorname*{null}\left(W_{\mathrm F}\right) \beta
	\end{equation}
	where we use $\beta$ as the decision variable instead of higher dimensional variable $\alpha$. Here, $\operatorname*{null}(\cdot)$ is the orthonormal basis of the null-space, and $\dagger$ denotes the pseudoinverse. Note that this substitution is possible because the disturbance forecast $\mathbf{w}_{[t:t+L-1]}$ is assumed to be exactly known in advance.
\end{rem}

\subsection{Connection to topology identification}
Indeed, there is a close conceptual relation between  approaches from topology and parameter estimation \cite{Chen2014, Yuan2023,Ardakanian2019} and our setting \eqref{eqn:dd_simplified}.

In the  power systems literature, the product $M = \Tilde{\mathbf{B}}^\mathrm{f} \Tilde{\mathbf B}^{-1}$ is known as the power transfer distribution factor (PTDF) matrix. 
Its $(i,k)$-th element describes the change in power flow through line $i \in \lbrace 1, \ldots, \abs{\mathcal{E}} \rbrace$ per unit increase in injected power at node $k$ \cite{Stott2009}. 
Following \cite{Yuan2023,Ardakanian2019}, the admittance matrix $\mathbf{Y}$ can be computed from the nodal voltage and current phasors by solving a least-squares problem.
Similarly, in the DC power flow setting, one can get an estimate $\hat{M}$ of the true PTDF matrix $M$ via least-squares
\begin{equation}
	\label{eqn:ptdf_caculation}
	\hat{M}^\star = \arg\min_{\hat{M}} \quad \norm{\mathscr{H}_1(\mathbf{f}^\mathrm{d}_{[1:T]}) - \hat{M} \cdot \mathscr{H}_1(\mathbf{P}^\mathrm{d}_{[1:T]})}_F^2,
\end{equation}
see, \cite{Chen2014}.
Here, signals $f^\mathrm{d}(t) \in \mathbb{R}^{\abs{\mathcal{N}}}$ and $P^\mathrm{d}(t) \in \mathbb{R}^{\abs{\mathcal{N}}}$ indicate the pre-recorded data, consisting of active power flows and nodal net power injections, respectively. 
Since the PTDF matrix has $\abs{\mathcal{N}}-1$ columns, a sufficient condition for its identification is $\rank \mathscr{H}_1(\mathbf{P}^\mathrm{d}_{[1:T]}) = \abs{\mathcal{N}}-1$, i.e., $\mathbf{P}^\mathrm{d}_{[1:T]}$ must be of full row rank. 
In practice, this condition might not be fulfilled due to the existence of nodes that accommodate neither generation nor demand, thus having zero net power injection for all $t \in \mathbb{N}_0$.

To reveal the connection between the data-based representation \eqref{eqn:dd_simplified} and the classical least-squares \eqref{eqn:ptdf_caculation}, first rewrite \eqref{eqn:dd_simplified} by eliminating $\alpha$. 
Neglecting the BESS dynamics for a moment, we obtain the following explicit predictor
\begin{equation*}
	\label{eqn:dd_predictor}
	\mathbf{y}_{[t:t+L-1]} =
	\tilde{M}
	\begin{bmatrix}
		\mathbf{u}_{[t:t+L-1]} \\
		\mathbf{w}_{[t:t+L-1]}
	\end{bmatrix}, \text{where }\tilde{M} = Y_{\mathrm F}
	\begin{bmatrix}
		U_{\mathrm F} \\
		W_{\mathrm F}
	\end{bmatrix}^\dagger.
\end{equation*}
The second equation above can then be rewritten in the least-squares form
\begin{equation*}
	\label{eqn:lin_regression}
	\tilde{M}^\star = \arg\min_{M} \quad \norm{Y_{\mathrm F} - \tilde{M} \begin{bmatrix}
			U_{\mathrm F} \\
			W_{\mathrm F}
		\end{bmatrix}}_F^2.
\end{equation*}
We consider $\tilde{M} = \begin{bmatrix}
		M_{U_\mathrm{F}} & M_{W_\mathrm{F}}
	\end{bmatrix}$ such that columns of $ M_{U_\mathrm{F}}$ and $M_{W_\mathrm{F}}$ correspond to rows of $U_{\mathrm F}$ and $W_{\mathrm F}$.
This yields
\begin{equation}
	\label{eqn:lin_regression1}
	M_{U_\mathrm{F},W_\mathrm{F}}^\star = \arg\min_{M_{U_\mathrm{F},W_\mathrm{F}}} \quad \norm{ Y_{\mathrm F} - M_{U_\mathrm{F}} U_{\mathrm F} - M_{W_\mathrm{F}} W_{\mathrm F}}_F^2.
\end{equation}
Next, in \eqref{eqn:dd_simplified} we set the prediction horizon to $L=1$ such that $U_\mathrm{F} = \mathscr{H}_1(\mathbf{u}^\mathrm{d}_{[1:T]})$, $W_\mathrm{F} = \mathscr{H}_1(\mathbf{w}^\mathrm{d}_{[1:T]})$. Assuming that only active power flows are measured, i.e., $y(t) \doteq f(t)$, we get $Y_\mathrm{F} = \mathscr{H}_1(\mathbf{f}^\mathrm{d}_{[1:T]})$. Since the net power injections $\mathscr{H}_1(\mathbf{P}^\mathrm{d}_{[1:T]})$ from \eqref{eqn:ptdf_caculation} are the differences between the total generated and consumed power at each node, we have that
\[
	\mathscr{H}_1(\mathbf{P}^\mathrm{d}_{[1:T]}) = \Tilde{\mathbf{C}}_\mathrm{g} U_{\mathrm F} - \tilde{\mathbf{C}}_\mathrm{d} W_{\mathrm F}.
\]
Substituting the above back into \eqref{eqn:ptdf_caculation} yields
\begin{equation}
	\label{eqn:lin_regression2}
	\hat{M}^\star\hspace{-.2em} =\hspace{-.0em} \arg \hspace{-.2em}\min_{\hat{M}}  \norm{ Y_{\mathrm F} - \hat{M}(\Tilde{\mathbf{C}}_\mathrm{g} U_{\mathrm F} - \tilde{\mathbf{C}}_\mathrm{d} W_{\mathrm F})}_F^2\hspace{-.4em}.\hspace{-.2em}
\end{equation}
Thus, the only difference between \eqref{eqn:lin_regression2} and the classic least squares \eqref{eqn:ptdf_caculation} is that the latter does not differentiate between specific generators and demands. At each time $t \in [1:T]$, all nodal active power injections are simply collected in the vector $P^\mathrm{d}(t)$ using the mappings $\tilde{\mathbf{C}}_\mathrm{g}$ and $\tilde{\mathbf{C}}_\mathrm{d}$.
Indeed, comparing \eqref{eqn:lin_regression2} and \eqref{eqn:lin_regression1} reveals that
\[
	M_{U_\mathrm{F}} = \hat{M} \Tilde{\mathbf{C}}_\mathrm{g}~\text{and}~M_{W_\mathrm{F}} = \hat{M} \Tilde{\mathbf{C}}_\mathrm{d},
\]
where $\hat{M}$ is the least-squares PTDF estimate from \eqref{eqn:ptdf_caculation}.
Hence, the topology estimation in \eqref{eqn:ptdf_caculation}, which is similar to \cite{Chen2014,Yuan2023,Ardakanian2019}, is implicitly included in Algorithm~\ref{alg:deepc} via \eqref{eqn:dd_sys}.

\section{Numerical Experiments}
We demonstrate the data-driven multi-stage OPF (DD OPF) on a modified IEEE 118-bus network with $\abs{\mathcal{G}} = 58$ (of which $\abs{\mathcal{S}} = 4$), $\abs{\mathcal{E}} = 186$, and $\abs{\mathcal{D}} = 99$. 
We compare our approach to: i) multi-stage OPF (MPC OPF) based on the exact PTDF model, and ii) classical sequential system identification and control (Scq ID) approach, where measured data is used first to calculate a PTDF estimate via \eqref{eqn:ptdf_caculation}.

To model the power system, we use the AC power flow calculation from Matpower \cite{Zimmerman2011}; the test network is taken from the Matpower database. 
Numerical optimization problems are solved using IPOPT 3.14.11 \cite{Waechter2006} interfaced through CasADi 3.6.7 \cite{Andersson2019}. 
The calculations are performed on a machine with an Intel Core i5-1235U processor and $16 \, \si{\giga \byte}$ RAM. 
The time step for the simulations is $\delta \doteq \num{0.25} \, \si{\hour}$. 
To sufficiently excite the data-driven controller we feed it with the solution of the multi-stage DC OPF. 
In practice, one could simply apply random disturbances to the already existing control policy instead.

In our modified IEEE 118-bus system, we add the BESS at nodes $\left\{21, 59, 89, 116\right\} \subset \mathcal{N}$. 
For each BESS, the minimum and maximum state of charge is set to $e^\mathrm{min} \doteq \num{0}$~MW$\unit{\hour}$ and $e^\mathrm{max} \doteq \num{200}$~MW$\unit{\hour}$, respectively; the power capability is constrained to $s^\mathrm{max} = -s^\mathrm{min} \doteq \num{50}$~MW. 
For the stage costs $\ell$, we use a quadratic form $\ell(u,y) \doteq u^\top \mathbf{R} u + \mathbf{r}^\top u + y^\top \mathbf{Q} y + \mathbf{q}^\top y$, with $\mathbf{R} \succ 0$ and $\mathbf{Q} \succeq 0$; the coefficients of the cost matrices are obtained from the Matpower database. 
For the power generation of each added BESS, we consider a quadratic cost of $\num{0.01}~\frac{\$}{\mathrm{MW}^2 \unit{\hour}}$. For the line power flows and stored energy, we set the quadratic cost to $\num{10}^{-5}~\frac{\$}{\mathrm{MW}^2 \unit{\hour}}$ ($\frac{\$}{\mathrm{MW}^2 \unit{\hour}^3}$). 
All remaining cost coefficients are set to zero. 
The line power flows are constrained to $f^\mathrm{max} = -f^\mathrm{min} \doteq \num{300}$~MW for all lines in $\mathcal{E}$.

To improve the computational performance of DD OPF, we reformulate the problem \eqref{eqn:dd_ocp} in terms of a segmented formulation \cite{o_dwyer.2023} with $L$ segments, each segment predicting $T_\mathrm{ini}=1$ time steps into the future, and thus containing Hankel matrices with $2 T_\mathrm{ini} = 2$ block rows (for the $T_\mathrm{ini}$-long initial and $T_\mathrm{ini}$-long future trajectory). 
This increases the number of constraints but allows to decrease the number of rows in the Hankel matrices. 
Additionally, relying on the result from \cite[p.~327]{Willems2005}, we know that the dimensionality of the input-disturbance-output behavior of the system \eqref{eqn:dc_pf_lds} is equal to $2 T_\mathrm{ini} (n_\mathrm{u} + n_\mathrm{w}) + \bm{q} = 2(\abs{\mathcal{G}} + \abs{\mathcal{D}} - 1) +  \abs{\mathcal{S}}$. 
This allows to decrease the number of columns in the Hankel matrices. 
Thus, we perform a low rank approximation via truncated singular value decomposition to ensure that the rank of the combined Hankel matrix is consistent with the dimension of the input-output behavior of the system. 
A similar procedure was documented in \cite{Coulson2019}.

First, in Table~\ref{tab:data_requirements} we compare the data requirements of the Scq ID approach with those of the presented DD OPF controller. 
DD OPF learns the BESS dynamics and does not require the knowledge of the demand and generator/BESS locations, which, however, comes at a cost of larger Hankel matrices. 
Note that the minimum data requirements of DD OPF do not depend on the number of buses $\abs{\mathcal{N}}$.

To make our closed-loop control simulations more realistic, we use additive Gaussian measurement noise to corrupt the measured line active power flows $\mathbf{f}^\mathrm{d}_{[1:T]}$ that we store in the Hankel matrices for both Scq ID and DD OPF. 
The noise-to-signal ratio is set to $1\%$.
\begin{figure*}[t]
	\centering
%
%
\definecolor{mycolor1}{rgb}{0.00000,0.44700,0.74100}%
\definecolor{mycolor2}{rgb}{0.85000,0.32500,0.09800}%
\definecolor{mycolor3}{rgb}{0.92900,0.69400,0.12500}%
\begin{tikzpicture}

\begin{axis}[%
width=0.9\textwidth,
height=4cm,
scale only axis,
xmin=1,
xmax=97,
ymin=0,
ymax=320,
xlabel={Time (15-minute intervals)},
ylabel={$\max_{i \in \mathcal{E}} \abs{f_i(t)}$ (MW)},
axis background/.style={fill=white},
legend style={at={(0.03,0.03)}, anchor=south west, legend cell align=left, align=left, draw=white!15!black}
]
\addplot[const plot, color=mycolor1] table[row sep=crcr] {%
1	269.534637750838\\
2	272.802259417332\\
3	275.791430247161\\
4	276.970303363057\\
5	288.659122753558\\
6	280.484489196542\\
7	274.24731331821\\
8	276.684495675383\\
9	268.714742204293\\
10	271.863026905715\\
11	277.176233551488\\
12	273.834464481231\\
13	276.251880455238\\
14	271.100636342793\\
15	269.342187850285\\
16	278.036138234296\\
17	288.989673147641\\
18	284.072528598191\\
19	297.070713655344\\
20	285.669754824093\\
21	286.5900016289\\
22	292.522974697941\\
23	297.893492330513\\
24	297.893529367594\\
25	297.893529564109\\
26	297.893529546208\\
27	297.893529502544\\
28	297.893529284234\\
29	297.893529498946\\
30	297.893529516126\\
31	297.893529421275\\
32	297.89352922491\\
33	297.893528024075\\
34	297.893523611648\\
35	297.893529462872\\
36	297.893529470705\\
37	297.893529098393\\
38	297.893527002563\\
39	282.604789646902\\
40	272.683656095032\\
41	262.39173054422\\
42	265.411423975089\\
43	268.147588917443\\
44	260.922302594782\\
45	264.034349981456\\
46	266.526633517748\\
47	270.765364698747\\
48	277.700743359708\\
49	269.474908280855\\
50	276.277431517517\\
51	265.950108160344\\
52	257.109691047098\\
53	295.89984910835\\
54	301.836994754043\\
55	266.30615175894\\
56	256.195863184756\\
57	272.230322957023\\
58	263.771053081915\\
59	262.17930373308\\
60	252.206599586625\\
61	253.851019561779\\
62	300.309604113987\\
63	300.309556416205\\
64	300.309594334728\\
65	300.309970006451\\
66	300.309696160123\\
67	300.309761478359\\
68	300.310578025968\\
69	300.309977545742\\
70	300.309675794777\\
71	300.309701200276\\
72	300.309570344345\\
73	300.309586220108\\
74	272.628610905895\\
75	258.700621844327\\
76	257.669863254484\\
77	253.561461484542\\
78	252.72025417288\\
79	248.096116859366\\
80	246.018551625866\\
81	244.453411099007\\
82	245.440732032299\\
83	246.746454287488\\
84	249.871571135365\\
85	242.739298667293\\
86	241.066134586699\\
87	224.487909296586\\
88	219.857029043447\\
89	230.898241416884\\
90	230.542518752088\\
91	247.88351905759\\
92	249.84621098412\\
93	254.523697436777\\
94	260.573124859982\\
95	257.429588477744\\
96	261.345532023365\\
97	261.345532023365\\
};
\addlegendentry{Exact DC model}

\addplot[const plot, color=mycolor2] table[row sep=crcr] {%
1	269.552265618083\\
2	272.81930756079\\
3	275.808719190955\\
4	276.985571004127\\
5	288.67496636137\\
6	280.500885640194\\
7	274.264174529246\\
8	276.702491659753\\
9	268.732274922808\\
10	271.881126351847\\
11	277.194051737416\\
12	273.852843768458\\
13	276.27047747759\\
14	271.118217221487\\
15	269.354025698873\\
16	278.04299155092\\
17	288.99559045138\\
18	284.067424362243\\
19	297.061698110046\\
20	285.661715614061\\
21	286.579510265137\\
22	292.511096002664\\
23	297.925436180581\\
24	299.153509338661\\
25	299.503881250831\\
26	298.942784875078\\
27	298.392877031079\\
28	299.050626786425\\
29	299.1705236807\\
30	298.346718150938\\
31	298.863469366463\\
32	298.290916834972\\
33	297.903616243332\\
34	298.559164313048\\
35	298.471838306886\\
36	298.243819091163\\
37	298.666086584651\\
38	298.289750881528\\
39	282.618696163327\\
40	272.697730178356\\
41	262.405782793\\
42	265.425142572338\\
43	268.160471277518\\
44	260.935706156615\\
45	264.04612694919\\
46	266.538661122853\\
47	270.777586438423\\
48	277.713827577579\\
49	269.488584657121\\
50	276.291004823635\\
51	265.96277448606\\
52	257.100174016317\\
53	295.467809767032\\
54	303.835329042719\\
55	266.296569889905\\
56	256.403184684771\\
57	272.048556249475\\
58	264.062956741256\\
59	262.73162133096\\
60	252.221077075667\\
61	253.865093433791\\
62	296.217952682893\\
63	296.626808008502\\
64	296.698235508345\\
65	297.396395258639\\
66	297.709670664885\\
67	298.130184780425\\
68	297.684300942766\\
69	297.590704207155\\
70	297.328456052146\\
71	297.218091401188\\
72	296.118330518415\\
73	296.023877124928\\
74	272.628367156467\\
75	258.497755131283\\
76	257.466802454137\\
77	253.359276755972\\
78	252.517743874773\\
79	247.891864152332\\
80	245.815806930761\\
81	244.251449201888\\
82	245.238798382055\\
83	246.544431861458\\
84	249.669887112965\\
85	242.753506357484\\
86	241.082185006309\\
87	224.503780329157\\
88	219.873359243303\\
89	230.914733174043\\
90	230.558269081057\\
91	247.899568134258\\
92	249.862369539235\\
93	254.540503784954\\
94	260.589316198008\\
95	257.446874622322\\
96	261.363983693279\\
97	261.363983693279\\
};
\addlegendentry{Scd ID}

\addplot[const plot, color=mycolor3] table[row sep=crcr] {%
1	203.695207868059\\
2	220.893980699602\\
3	241.908424991319\\
4	204.099162965394\\
5	224.208282430905\\
6	246.362495890041\\
7	244.64407612633\\
8	260.734884358024\\
9	255.474026428221\\
10	292.824712987913\\
11	305.97187868676\\
12	290.902206767946\\
13	283.748692007431\\
14	271.229029877762\\
15	244.582680086982\\
16	221.899405261445\\
17	201.563772064168\\
18	198.384495243156\\
19	206.501811891563\\
20	187.866561572969\\
21	188.217119150861\\
22	180.995140324557\\
23	189.061067615049\\
24	204.130606511164\\
25	210.83172639806\\
26	237.378036466975\\
27	249.34372316947\\
28	230.254555707236\\
29	224.127682065902\\
30	207.83351427664\\
31	193.593425226919\\
32	193.250356262586\\
33	206.24860337045\\
34	201.287452280664\\
35	235.340329012003\\
36	208.230087223017\\
37	218.352482825275\\
38	184.041197579649\\
39	204.690009403111\\
40	218.671726035838\\
41	169.629150543596\\
42	168.41983907137\\
43	151.992253380303\\
44	161.826141908812\\
45	196.569905300776\\
46	203.837265434008\\
47	205.645284000107\\
48	198.34104906548\\
49	194.796378106117\\
50	194.892545247822\\
51	192.266685276316\\
52	196.607914700279\\
53	240.815935290632\\
54	246.021413516009\\
55	209.058630303662\\
56	206.633325108533\\
57	223.636556228494\\
58	196.018361915979\\
59	186.880912859501\\
60	178.996331443893\\
61	210.724977315939\\
62	295.360549910388\\
63	303.016350842904\\
64	302.514282283658\\
65	303.767272329098\\
66	304.226600120392\\
67	304.033449217956\\
68	302.579756796665\\
69	303.738670980936\\
70	303.47464755337\\
71	303.221347788467\\
72	303.261118733397\\
73	302.710474761053\\
74	273.293263525434\\
75	239.286705718416\\
76	232.203929087838\\
77	189.125694393025\\
78	161.136189708958\\
79	137.91664909274\\
80	146.912813820471\\
81	143.369771452127\\
82	139.767034380072\\
83	147.027960839669\\
84	157.749898200431\\
85	164.482756031248\\
86	206.030429725493\\
87	181.530512439116\\
88	182.9298663216\\
89	185.303093028132\\
90	178.863514935858\\
91	196.68951461481\\
92	178.843828179581\\
93	189.847671295356\\
94	188.707068214992\\
95	191.113223644508\\
96	184.338142026467\\
97	184.338142026467\\
};
\addlegendentry{DD OPF}

\addplot [color=black, forget plot]
  table[row sep=crcr]{%
1	300\\
97	300\\
};
\addlegendentry{Upper bound}

\end{axis}
\end{tikzpicture}%
	\caption{Maximum power flow over all lines during closed-loop control for the three considered controllers.}
	\label{fig:max_power_flows}
\end{figure*}
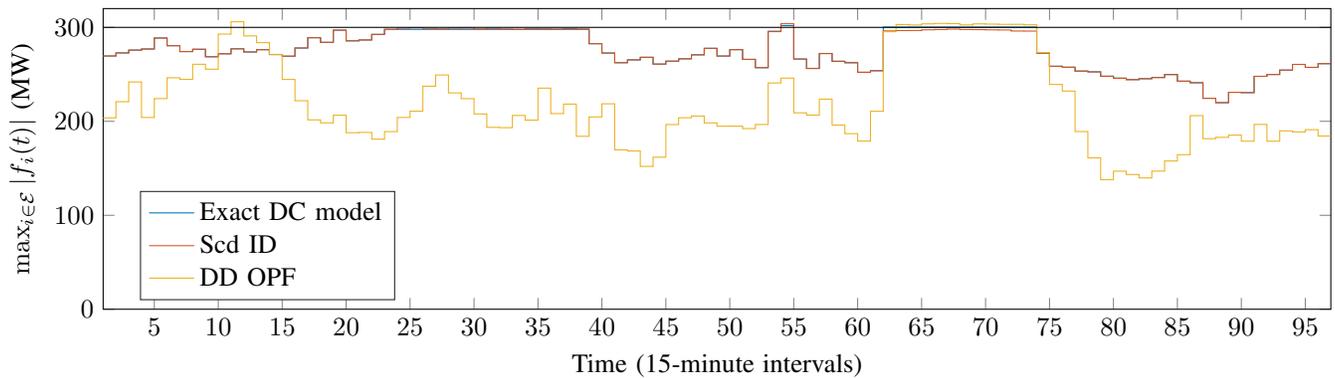
It is, however, worth noting that the equation \eqref{eqn:dd_simplified} stemming from the Willems' lemma only holds for noise-free data. 
Hence, for DD OPF we add a quadratic regularization term $\lambda \norm{\beta}_2^2$ with $\lambda \doteq \num{200}$ for $\beta$ from \eqref{eqn:substitution} to the objective function \eqref{eqn:dd_costs} to avoid overfitting.

In Table~\ref{tab:my-table}, we compare the performance of the classic multi-stage DC OPF \eqref{eqn:dc_pf_mpc} based on exact topology data $(\mathbf{B}, \mathbf{B}_\mathrm{f})$, the Scq ID approach, and the presented direct data-driven control \eqref{eqn:dd_ocp} with trajectory segmentation. 
The OCPs are solved for a prediction horizon of $L \doteq 12$ for a randomly-generated demand forecast. The solutions are applied in closed-loop in a receding horizon fashion, with control horizon $L_c=1$ ( cf.~Algorithm~\ref{alg:deepc}). 
This control is performed for $\num{96}$ time intervals (1 day). To learn the dynamics \eqref{eqn:dc_pf_lds}, we generate a pre-recorded trajectory $(\mathbf{u}^{\mathrm d}, \mathbf{w}^{\mathrm d}, \mathbf{y}^{\mathrm d})_{[1:T]}$ of length $T=417$. 
This data is used by both Scq ID and DD OPF.

Table~\ref{tab:my-table} summarizes the realized closed-loop control costs $J_\mathrm{CL}$ and the median OCP solution time (among all of the $\num{96}$ OCPs). 
The DD OPF controller does not require the knowledge of generator, demand, or BESS locations, while additionally learning the dynamics of the BESS. 
Yet, the realized closed-loop costs are only $0.5\%$ higher than the costs incurred from using the classical, sequential identification and control approach, whereas the OCP solution time is satisfactory considering the chosen time scale.

Figure~\ref{fig:max_power_flows} shows the maximum active power flow over all lines $\mathcal{E}$ of the system over time during closed-loop control. 
The results suggest that using DD OPF yields control inputs different from those of the classical approaches such as multi-stage OPF based on exact PTDF matrix and Scq ID. 
This is due to the bias introduced by the quadratic regularization of $\beta$ in \eqref{eqn:substitution}. 
Nevertheless, as shown in Figure~\ref{fig:max_power_flows}, the resulting power flows are still kept within the upper bound of 300~MW, except for a few time steps.

\begin{table}
	\centering
	\caption{Minimum data requirements for the considered controllers}
	\label{tab:data_requirements}
	\begin{tabular}{l|l}
		Scq ID                                                                  & DD OPF                                                                                \\ \hline \hline
		Locations of demands and                                                & Combined Hankel matrix                                                                \\
		generators/BESS, data matrix                                            & $(\abs{\mathcal G} + \abs{\mathcal E} + \abs{\mathcal D} + 3\abs{\mathcal S}) \times$ \\
		$(\abs{\mathcal E} + \abs{\mathcal N} - 1) \times (\abs{\mathcal N}-1)$ & $(2(\abs{\mathcal G} + \abs{\mathcal D} - 1) + \abs{\mathcal S})$                     \\ \hline
	\end{tabular}
\end{table}

\begin{table}
	\centering
	\caption{Comparison of the controllers in the 118-bus system}
	\label{tab:my-table}
	\begin{tabular}{l|lll}
		\begin{tabular}[c]{@{}l@{}}Performance\\ metric\end{tabular} & \multicolumn{1}{l|}{\begin{tabular}[c]{@{}l@{}}Exact\\ DC model\end{tabular}} & \multicolumn{1}{l|}{Scq ID} & DD OPF \\ \hline \hline
		$J_\mathrm{CL}${[}k\${]}                                     & \multicolumn{1}{l|}{1.6978}                                                   & \multicolumn{1}{l|}{1.6979} & 1.7065 \\ \hline
		$t_\mathrm{OCP}$ [s] (median)                                & \multicolumn{1}{l|}{0.82}                                                     & \multicolumn{1}{l|}{0.91}   & 5.09   \\ \hline
	\end{tabular}
\end{table}

\section{Conclusion and Outlook}
This paper has explored a data-driven approach to multi-stage DC OPF. 
The proposed approach transfers concepts from data-driven control to power systems. 
We have shown that our one-shot approach entails a classic sequential approach and does not need any topology information about the considered grid. 
Instead, only generation and demand time series are required. 
Future work will explore the consideration of measurement noise and the extension towards voltage data.

\bibliographystyle{ieeetr}
\bibliography{literature}

\end{document}